\documentclass[a4paper]{article}
\usepackage[USenglish]{babel} %francais, polish, spanish, ...
\usepackage[T1]{fontenc}
\usepackage[ansinew]{inputenc}

\usepackage{lmodern} %Type1-font for non-english texts and characters

\usepackage{graphicx} %%For loading graphic files
\usepackage{amsmath}
\usepackage{amsthm}
\usepackage{amsfonts}

\begin{document}

\pagestyle{empty} %No headings for the first pages.

%% Title Page %%%%%%%%%%%%%%%%%%%%%%%%%%%%%%%%%%%%%%%%%%%%%%%
%% ==> Write your text here or include other files.

%% The simple version:
\title{Renormalization in  an interpolating gauge in Yang-Mills theory}
%\author{A Andrasi}

\author{A Andra\v si$^+$ and J C Taylor\footnote{Corresponding author } 
\footnote{\textit{E-mail addresses} aandrasi@irb.hr (A. Andrasi), jct@damtp.cam.ac.uk (J. C. Taylor)} \\ \\ {\it  $^+$Vla\v ska 58, Zagreb, Croatia} \\ $^\dagger${\it DAMTP, Cambridge University, UK}}
%\date{} %%If commented, the current date is used.
%\affiliation{$^a$ Vla\v ska 58, Zagreb, Croatia\footnote[]{August 30, 2018}}
%\affiliation{$^b$ DAMTP, Cambridge University, UK}

%% The nice version:
%\input{titlepage} %%You need a file 'titlepage.tex' for this.
%% ==> TeXnicCenter supplies a possible titlepage file
%% ==> with its templates (File | New from Template...).

%% Inhaltsverzeichnis %%%%%%%%%%%%%%%%%%%%%%%%%%%%%%%%%%%%%%%
%\tableofcontents %Table of contents
%\cleardoublepage %The first chapter should start on an odd page.

%% Chapters %%%%%%%%%%%%%%%%%%%%%%%%%%%%%%%%%%%%%%%%%%%%%%%%%
%% ==> Write your text here or include other files.

%\input{intro} %You need a file 'intro.tex' for this.

\maketitle
%%%%%%%%%%%%%%%%%%%%%%%%%%%%%%%%%%%%%%%%%%%%%%%%%%%%%%%%%%%%%

%% ==> Some hints are following:

%\chapter{Some small hints}\label{hints}

\begin{abstract}

\noindent{The Coulomb gauge in QCD is the only explicitly unitary gauge. But it suffers from energy-divergnces which means that it is not rigorously well-defined. One way to define it unambiguously is as the limit of a gauge interpolating between the Feynman gauge and the Coulomb gauge. This interpolating gauge is characterized by a parameter $\theta$ and the Coulomb gauge is obtained in the limit $\theta \rightarrow 0$. We study the renormalization of this $\theta$-gauge for all values of $\theta$.
Novel features include field mixing as well as scaling,  the renormalization of the $\theta$ parameter itself, and the appearance of new counter-term structures at two-loop order.}\\

\noindent{Pacs numbers: 11.15.Bt; 03.70.+k}\\

\noindent{Keywords QCD, Coulomb gauge, renormalizatio}

\end{abstract}

\vfill\newpage

\pagestyle{plain} %Now display headings: headings / fancy / ...

\section{Introduction}
The Coulomb gauge in non-abelian gauge theory is the only explicitly unitary gauge.
But in perturbation theory it suffers from 'energy divergences', that is Feynman integrals which are divergent  over the
time-components of the momenta, while the spacial components are held fixed.
The simplest such energy divergences occur at one loop. For pure YM theory, these are quite easily canceled by combining Feynman
diagrams appropriately, but they are automatically removed by using the Hamiltonian, rather than the Lagrangian, formalism \cite{mohapatra}.
%When quark loops are included, Ward identities secure the cancellation of this type of energy-divergence \cite{aajct1}.

More subtle, logarithmic, divergences appear first at two loop order. The cancellation of these was proved by Doust \cite{doust} (see also \cite{cheng}
and generalized in \cite{aajct2}).
The origin of these divergences was linked by Christ and Lee \cite{christlee} with the problem of correctly ordering the factors in the
Coulomb potential in the Hamiltonian (see also \cite{schwinger}). But in this paper we consider only ordinary momentum-space Feynman perturbation theory,
in the manner of Doust.

 Pure energy-divergences, that is divergences over the energy integrals with all spatial momenta fixed, occur
at 2-loop order only, not at higher order (see \cite{doust}). But if ordinary UV divergences are combined with energy divergences, new problems occur at 3-loop order, from the insertion of  UV divergent  loops into
two-loop  graphs.
A difficulty in studying this is to be sure that the divergent integrals we are manipulating are well-defined.
To overcome this problem, we make use of a 'flow gauge', which interpolates between the Feynman gauge
and the Coulomb gauge. This flow gauge is characterized by a parameter $\theta$, $\theta=1$ is the Feynman
gauge, and the  Coulomb gauge is defined by the limit $\theta \rightarrow 0$. For nonzero $\theta$, there are no energy divergences
in any Feynman integral. We have used this flow gauge in a previous paper \cite{aajct1} to study the insertion of UV divergent loops into energy-divergent graphs.
 We emphasize that the flow gauge is of no practical use, having  the advantages of neither the Feynman nor the Coulomb gauge.
We use it only as a mathematical tool.

To 2-loop order, Doust \cite{doust} has shown that the summed energy divergences give the $O(\hbar^2)$ terms in the Hamiltonian which were derived by Christ and Lee by consideration of operator ordering.  The 3-loop energy divergences  give higher order corrections, but they are not energy-independent like the  Christ-Lee operator. The main purpose of the present paper is to complete the renormalization of the flow gauge for general values of $\theta$. It turns out that the renormalization is complicated by three features: (i) there is mixing between fields (and sources) as well as scaling; (ii) there are new counter-term structures which appear for the first time at two-loop order, which are not needed to one-loop order; (iii) the gauge parameter $\theta$ itself is scaled in renormalization. 

In section 2, we review the flow gauge. In section 3, we discuss the renormalization to one-loop order, and give the values of the counter-terms for general values of $\theta$. There is a counter-term ($a_{12}$ in equation (24) below) which is not needed for $\theta=0$. In section 4, we give the structure of the counter-term for general loop order
 $n$. In section 5, we indicate how renormalization could be done by iteration from loop order $n$ to
 $n+1$.

\section{The interpolating gauge}
We use indices $i,j,k;m,n$ for spatial vectors; $\lambda, \mu, \nu$ for Lorentz vectors, $a,b,c,d$ for colour. We generally suppress colour indices, and use the notation
\begin{equation}
(X\wedge Y)^a=f^{abc}X^bY^c.
\end{equation}
%1
Energy divergences occur when there are integrals which are divergent over the energy variables, with the spatial momenta held fixed.
These divergences are removed by going to a gauge defined by the gauge-fixing term
\begin{equation}
L_{GF}=-\frac{1}{2\theta^2}[(\partial_i A^i+\theta^2 \partial_0 A^0)^2].
\end{equation}
%2
 For $\theta=1$ this gauge is the Feynman gauge, and the limit
$\theta \rightarrow 0$ gives the Coulomb gauge by enforcing $\partial_iA^i =0$. 

The Yang-Mills Lagrangian in the Hamiltonian formalism is
\def\t{\theta}
\begin {equation}
L_0= -\frac{1}{4}F_{ij}.F^{ij} +\frac{1}{2}E_i.E^i-E^iF_{0i}$$
$$+u^i.[\partial_ic+g(A_i\wedge c)]+u_0[\partial_0c+g(A_0\wedge c)]+gv^i.(E_i\wedge c)
-\frac{1}{2}gK.(c\wedge c)$$
$$+\partial_ic^*.[\partial^ic+gA^i\wedge c)]
+\t^2\partial_0c^*(\partial_0c+gA_0 \wedge c)
\end{equation}
%3
Here
\begin{equation}
F_{\mu\nu}=\partial_{\mu}A_{\nu}-\partial_{\nu}A_{\mu}+g(A_{\mu}\wedge A_{\nu}),
\end{equation}
%4
$E_i^a$ is the momentum conjugate field to $A_i^a$ (that is the colour electric field), $c$ is the ghost, $c^*$ the anti-ghost, and  $u_i, u_0, v_i, K$ are the sources for the gauge transforms of
$A_i, A_0, E_i, c$ respectively.

 We will use  the notation that the momentum $k=(k_0,\textbf{K})$, K=|\textbf{K}|, and $k^2=k_0^2-K^2$, and
\def\K{\bar{K}}
\begin{equation}
\K^2\equiv \textbf{K}^2-\theta^2 k_0^2.
\end{equation}
%5
\def\c{\gamma}
\def\t{\theta}
\def\R{\bar{R}}
\def\P{\bar{P}}
\def\Q{\bar{Q}}
We use $g_{\mu\nu}$ for the Minkowski metric with $g_{00}=1$. In the gauge given by (2), the Coulomb propagator is
\begin{equation}
-\frac{1}{\K^2},
\end{equation}
and the ghost propagator by
\begin{equation}
  +\frac{1}{\K^2}
	\end{equation}
%7
the spatial propagator is
\begin{equation}
\frac{1}{k^2+i\eta}\left[g_{ij}+(1-\t^2)\frac{K_i K_j}{\K^2}\right]=\frac{1}{k^2+i\eta}\left[g_{ij}+\frac{K_i K_j}{K^2}\right]+\t^2\frac{K_iK_j}{K^2\K^2},
\end{equation}
%8
where the second form displays the transverse and longitudinal  parts.
Since we use the Hamiltonian formalism, we require also propagators involving the electric field $\bf{E}$. It is
\begin{equation}
\frac{K^2}{k^2+i\eta}\left[ g^{mn} +\frac{K^m K^n}{\textbf{K}^2} \right].
\end{equation}
%9
(unlike (8), this is transverse.) (We use indices $i,j,...$ for the potential $\bf{A}$ and indices $m,n,...$ for $\bf{E}$).
There are also off-diagonal propagators. That between $E^m$ and $A_j$ is
\begin{equation}
\frac{ik_0}{k^2+i\eta}\left[ g^m_j+(1-\t^2)\frac{K^mK_j}{\K^2} \right],
\end{equation}
%10
and the off-diagonal propagator between $E_m$ and $A_0$ is
\begin{equation}
\frac{iK^m}{\K^2}.
\end{equation}
%11?
In addition to the usual Yang-Mills vertex involving the spatial components $A_i$, there is an 
$E^{ma}A_0^bA_i^c$ vertex
\begin{equation}
-gf^{abc}g^i_m.
\end{equation}
%12
With the gauge-fixing term (2), there is no off-diagonal propagator between $A_i$ and $A_0$.
%\begin{figure}[t]
	%\centering
		%\includegraphics[width=0.5\textwidth]{Fig1a.eps}
	%\label{fig:Fig1a.eps}
%\end{figure}
%\begin{figure}[h]
	%\centering
		%\includegraphics[width=0.5\textwidth]{Fig1b.eps}
	%\label{fig:Fig1b}
%\end{figure}
%\begin{figure}[h]
%\centering
%\includegraphics[width=0.8\textwidth]{Fig1a.eps}
%\end{figure}
\section{Renormalization to one-loop order}
In order to renormalize the flow gauge we need to use BRST invariance (see for example \cite{IZ} section 12.4).
We denote the BRST operation by $*$, where for any two space-time integrals $X,Y$
\begin{equation}
X*Y \equiv \int d^4x\left[\frac{\delta X}{\delta u^i(x)}.\frac{\delta Y}{\delta A_i(x)}+\frac{\delta X}{\delta u^0(x)}.\frac{\delta Y}{\delta A_0(x)}+\frac{\delta X}{\delta v^i(x)}.\frac{\delta Y}{\delta E_i(x)}+\frac{\delta X}{\delta c(x)}.\frac{\delta Y}{\delta K(x)}\right]$$
$$\pm (X\leftrightarrow Y),
\end{equation}
where the plus sign applies if at least one of $X,Y$ is bosonic, and the minus sign applies if they are both fermionic (that is, anti-commuting).
%13

We define
\begin{equation}
\gamma_0= \int d^4xL_0(x).
\end{equation}
%14
We use $\gamma$ for the renormalized and unrenormalized actions, and $\Gamma$ for the complete effective action derived from (14). $\gamma_0$ satisfies $\gamma_0 * \gamma_0 =0$, and as a consequence $\Gamma$ has the same property (see for example \cite{IZ} section (12-4-2)):
\begin{equation}
\Gamma * \Gamma =0,
\end{equation}
%15
and the additional relation
\begin{equation}
\frac{\delta \Gamma}{\delta c^*}=-\partial_i\frac{\delta\Gamma}{\delta u_i}-\t^2\partial_0\frac{\delta\Gamma}{\delta u_0}
\end{equation}
%16
which determines the dependence on $c^*$  and on $\t$.
If, in (13),   $Y$ is fermionic (that is, anti-commuting) and $X$ is bosonic (or the other way round), the order of the factors  is important. For example then $\delta X/\delta u^i$ and $\delta Y/\delta A_i$ do not commute. The order shown in (13) ensures that it implies the idempotent condition 
\begin{equation}
\Gamma * \{\Gamma * G\}=-\frac{1}{2}[\{\Gamma * \Gamma\} *G]=0
\end{equation}
%17
for any $G$ (bosonic or fermionic).

We will use the notation
\begin{equation}
\Gamma^{(N)} =\sum_0^N \Gamma_n 
\end{equation}
%18
for the sum of the contributions up to $N$ loops from (that is the contributions up to order $\hbar^N$).
Then (15) implies that (note that $\gamma_0$ is the same as $\Gamma_0$)
\begin{equation}
\gamma_0 * \Gamma_1 =0.
\end{equation}
%19
The UV divergent part of $\Gamma_1$ satisfies the same equation and hence also the counter-terms
(to 1-loop order), $\gamma_1$,
satisfy
\begin{equation}
\gamma_0 * \gamma_1=0.
\end{equation}
%20
Using eq (17), solutions of this equation are generated by
\begin{equation}
\gamma_1=\gamma_0 * G_1.
\end{equation}
%21
The most general form of $G_1$ allowed by locality, dimensions, rotational invariance, ghost number, and
invariance under the rigid colour group, is
\begin{equation}
G_1=a_5(u^i+\partial^ic^*).A_i+a_6(u_0+\t^2\partial_0c^*).A_0+a_7c.K+a_8v^i.E_i$$
$$+v^i.\left[a_9\partial_iA_0+a_{10}\partial_0A_i+a_{11}A_0 \wedge A_i+ a_{12} \frac{1}{2}(v_i \wedge c)\right],
\end{equation}
%22
where the $a_n$  are arbitrary divergent constants (proportional to $1/(4-d)$ in dimensional regularization). The $c^*$ terms are inserted in order to be consistent with (16). The numbering is to accord with our conventions in \cite{aajct4} ($a_{12}$ was not necessary
to zero order in $\theta$.)

In addition to counter-terms of the form (21), there is
\begin{equation}
a_0g\frac{d\gamma_0}{dg}
\end{equation}
%23
which also satisfies (20), as can be seen by differentiating $\gamma_0 *\gamma_0 =0$ with respect to $g$.

\def\w{\wedge}
\def\p{\partial}
The counter-terms from (21) and (23),  are
\begin{equation}
-a_5(\p_iA_j-\p_jA_i).\p^iA^j-(3a_5+a_0)g\p_iA_j.(A^i\w A^j)-(a_0/2+a_5)g^2(A_i\w A_j).(A^i\w A^j)$$
$$-[a_9\p_iA_0+a_{10}\p_0A_i+a_{11}(A_0\w A_i)].F^{0i}
$$
$$-(a_5+a_8-a_{10})E^i.\p_0A_i+(a_6+a_8+a_9)E^i.\p_iA_0-\{(a_5+a_6+a_0+a_8)g-a_{11}\}E^i.(A_0\w A_i)$$
$$ +a_8E_i.E^i
-(a_5+a_7)(u^i+\p^ic^*).\p_ic-(a_6+a_7)(u^0+\t^2\p_0c^*).\p_0c$$
$$-(a_9+a_{10})v^i.\p_i\p_0c
+(a_0-a_7)g[(u^i+\p^ic^*).(A_i\w c)+(u^0+\t^2\p_0c^*).(A_0 \w c)]+(a_0-a_7)gv^i.(E_i\w c)$$
$$+\frac{1}{2}(a_7-a_0)gK.(c\w c)
-(a_9g+a_{11})v^i.(A_0\w \p_ic)-(a_{10}g-a_{11})v^i.(A_i\w \p_0c)$$
$$-\frac{1}{2}a_{12}g(v^i\w c).(v_i\w c)+a_{12}(E^i-F^{0i}).(v_i\w c).
\end{equation}
%24
\def\z{(2\pi)^{-4}}
By calculating the UV divergent parts of  the amplitudes in (24), we have deduced the values of 
counter-terms. They are expressed in terms of the divergent constant
\begin{equation}
    c= \frac{g^2}{8\pi^2}C_G\frac{1}{4-d},
		\end{equation}
		%25
where $C_G$ is the colour group Casimir. The results are:
\newpage
\begin{equation}
a_0=-\frac{11}{6}c,\,\,\,\,\,
a_5=\left[\frac{1}{2}+\t-\frac{4}{3}\frac{\t^2}{1+\t}\right]c,\,\,\,\,
a_6=\left[\frac{11}{6}-\t\right]c,$$		
$$a_7=\left[-\frac{11}{6}+\frac{1}{2}\t\right]c,\,\,\,\,
a_8=\left[-\frac{2}{3}+\frac{1}{2}\t-\frac{2}{3}\frac{\t^2}{1+\t}\right]c,$$
$$a_9=\left[\frac{1}{6} -\frac{\t(1-\t)}{1+\t}\right]c,\,\,\,\,
a_{10}=\left[-\frac{1}{6}+\frac{2}{3}\frac{\t^2(1-\t)}{(1+\t)^2}\right]c,$$
$$a_{11}=g\left[-\frac{1}{6}-\frac{\t^3}{(1+\t)^2}\right]c,\,\,\,\,
a_{12}=-g\frac{1}{3}\frac{\t^2(2+\t)}{(1+\t)^2}c.
\end{equation}
%26
$a_0$ is independent of $\t$, and for $\t=1$ (the Feynman gauge) $a_5=a_6=\frac{5}{6}c$.
In  Appendix A, we give one example of the  calculation of a divergent part.

 Note that the renormalized action $\gamma_0+\gamma_1$  is no longer strictly  of Hamiltonian form: the equations of motion contain second order  time derivatives coming from the second line in (24).

\section{Renormalization to all orders}
In this section, we show how the work of the last section can be extended to
all loop orders, by a combination of field and source scaling and mixing.
We divide the effective action $\Gamma$ into two  parts, denoted by \textit{tilde} and \textit{hat},
\begin{equation}
\Gamma=\widetilde{\Gamma} +\widehat{\Gamma}
\end{equation}
%27
where $\widehat{\Gamma}$ is the part proportional to $c^*$ and $\widetilde{\Gamma}$ is the rest. 
If $\widetilde{\Gamma}$ is known, $\widehat{\Gamma}$ is determined by equation (16), that is
\begin{equation}
\frac{\delta \widehat{\Gamma}}{\delta c^*}=-\partial_i\frac{\delta \widetilde{\Gamma}}{\delta u_{i}}-\t^2\partial_0 \frac{\delta \widetilde{\Gamma}}{\delta u_{0}}.
\end{equation}
%28
Note that this equation also fixes the dependence on $\t$.

Define  renormalized fields and sources:
\begin{equation}
A_R^i=Z_{5}^{1/2}A^i,\,\,\,\,A_R^0=Z_{6}^{1/2}A_0,\,\,\,\,c_R=Z_7^{-1/2}c,\,\,\,\,v^i_R=Z_{8}^{-1/2}v^i,$$
$$u^i_R=Z_5^{-1/2}\left[u^i+Y_{10}\partial_0v^i-Y_{11}v^i\wedge A^0\right],$$
$$u_R^0=Z_{6}^{-1/2}\left[u_0+Y_{9}\partial_iv^i+Y_{11}v^i\wedge A_i\right],$$
$$E^i_R=Z_{8}^{1/2}\left[E^i+Y_{9}\partial^i A_0+Y_{10}\partial_0 A^i+Y_{11}A_0\wedge A^i+Y_{12}v^i\wedge c\right],$$
$$K_R=Z_{7}^{1/2}\left[K+\frac{1}{2}Y_{12}v^i\wedge v_i\right].
%29
\end{equation}
This represents mixing, given by the $Y$ coefficients, followed by scaling, given by the $Z$ coefficients. (We define the $Z$'s with square roots to follow the usual notation in QED and QCD.)	
These coefficients are related to the $a$'s in the last section by
\begin{equation}
Z_{5}^{1/2}=1 +a_5+......,
\end{equation}
%30
and similarly for $a_6,a_7,a_8$; and
\begin{equation}
Y_{9} = a_9+....,
\end{equation}
%31
and similarly for $a_{10},a_{11},a_{12}$.
We also define
\begin{equation}
g_R=Z_{0}^{1/2}g,
\end{equation}
%32
where $Z_0$ is related to $a_0$ similarly to equation (30).

With these definitions, we define the renormalized (that is, bare) action to be
\begin{equation}
\widetilde{\gamma}_R=\widetilde{\gamma}_0(A_R^i, A_R^0, E_R^i, c_R, u_R^i,u_R^0,v_R^i,K_R;g_R).
\end{equation}
Here the function $\widetilde{\gamma}_0$ is the part of (14) with terms in $c^*$ omitted, that is derived from the first two lines of $L_0$ in (3).
%33
Then we will show that 
\begin{equation}
\widetilde{\gamma}_R *\widetilde{\gamma}_R=0.
\end{equation}
%34

The renormalized quantities defined in (29) have the property that
\begin{equation}
A_{Ri}^a(x) * u^{bj}_R(y)=g^j_i\delta^{ab}\delta(x-y),\,\,\,A_{R0}^a(x) * u_{R0}^b(y)=\delta^{ab}\delta(x-y),$$ 
$$
E_{Ri}^a(x) * v_R^{jb}=g_i^j\delta^{ab}\delta(x-y),\,\,\,K_R^a(x) * c_R^b(y)=\delta^{ab}\delta(x-y),
\end{equation}
%35
and all the other 32 pairs of fields/sources give zero. For example
\begin{equation}
E_{Ri}^a(x) * u_{R0}^b(y)=0,\,\,\,\,\,E_{Ri}^a(x) * u_{Rj}^b(y)=0,$$
$$u_R^{ia}(x) * u_{R0}^b(y)=0,\,\,\,\,\, E_{Ri}^a(x) * K^b(y)=0.
\end{equation}
%36
These last four relations are not obvious, but depend on cancellations between different contributions in (13). Because of this property of the transformation (29), it is what Weinberg \cite{wein} calls 'anticanonical'. The required property (34) follows as a consequence.. In Appendix B we give a proof applicable to the present case.

Given now $\widetilde{\gamma}_R$ defined by (33) and satisfying (34), $\widehat{\gamma}_R$ is given by (28), that is
\begin{equation}
 \frac{\delta \widehat{\gamma}_R}{\delta c^*(x)}=-Z_5^{-1/2}\frac{\partial}{\partial x^i}\left[\frac{\delta \widetilde{\gamma}_R}{\delta u_{iR}(x)}\right]-Z_6^{-1/2}\t^2\frac{\partial}{\partial x_0}\left[\frac{\delta \widetilde{\gamma}_R}{\delta u_{0R}(x)}\right].
\end{equation}
%37
Inserting (33) and using the notation
\begin{equation}
c^*_R=Z_5^{1/2}c^*,\,\,\,\,\,\t_R^2=(Z_5/Z_6)^{1/2}\t^2,
\end{equation}
%38
we get
\begin{equation}
\widehat{\gamma}_R=\int dx [\partial_ic^*_R.(\partial^ic+g_RA^i_R\wedge c_R)
+\t^2_R\partial_0c^*_R.(\partial_0c+g_RA_{0R} \wedge c_R)],
\end{equation}
%39
which may be considered as the renormalized form of the last line of (3).
For the Feynman gauge, $\t=1$ we have that $\t_R=1$ also.

In terms of the renormalized quantities, the gauge-fixing term (2) becomes
\begin{equation}
L_{GF}=-\frac{1}{2\theta_R^2}(Z_5Z_6)^{-1/2}[(\partial_i A_R^i+\theta^2_R \partial_0 A_R^0)^2].
\end{equation}
%40
\newpage
$\widetilde{\gamma}_R$ in (33) contains a part linear in the $Y$ coefficients and a part quadratic in them.
The linear part is similar to (24), bearing in mind (30), (31) and (32). But the quadratic part 
\begin{equation}
\int dx\frac{1}{2}Z_8[Y_9\partial_iA_0+Y_{10}\partial_0A_i+Y_{11}A_0\wedge A_i+Y_{12}v_i\wedge c]$$
$$\times .[Y_9\partial^iA_0+Y_{10}\partial_0A^i+Y_{11}A_0\wedge A^i+Y_{12}v^i\wedge c]
\end{equation}
%41
has a different form from the second line of (24), and contributes first at 2-loop order.  
We give a simple example of terms coming from (41). Take counter-terms of the form
\begin{equation}
\alpha_1 \partial_0 A_i . \partial^i A_0 + \alpha_2 \partial_i A_0 .\partial^iA_0+\alpha_3 \partial_0 A_i .\partial_0 A^i,
\end{equation}
%42
and consider the combination
\begin{equation}
\alpha \equiv \alpha_1 +\alpha_2+\alpha_3.
\end{equation}
%43
In $\tilde{\gamma}_R$ in (33), terms linear in $Y_9$ and $Y_{10}$ from the fourth line of (29) contribute to (42); but in (43) they cancel out. In particular, $\alpha$ is zero to one-loop order. But (41) does contribute to (43) and gives
\begin{equation}
\alpha=\frac{1}{2}Z_8(Y_9+Y_{10})^2.
\end{equation}
%44
To two-loop order this reduces to
\begin{equation}
\alpha=\frac{1}{2}(a_9+a_{10})^2.
\end{equation}
%45
Thus the two-loop contribution to $\alpha$ is predicted just from a  knowledge of the one-loop
counter-terms in (26), and the corresponding divergent part is of course the negative of (45).
There are several similar examples involving $a_{11}$ and $a_{12}$.

\section{Renormalization order-by-order}
Renormalization is an iterative process. Renormalization to any given order, say $N$  loops, has to be completed before the next order, $N+1$ loops, is started. In this section, we show that this process is consistent with the complete renormalized action (33).

For the renormalized quantities up to a finite order $N$ of loops, we use a notation, illustrated by two examples,
\begin{equation}
[Z_8^{(N)}]^{1/2}=\sum_0^N a_{8,n},\,\,\,Y_9^{(N)} =\sum_1^N a_{9,n},
\end{equation}
%46
where as in (18) the superfix $N$ is the maximum loop order and the suffix $n$ is an individual contribution of order $\hbar^n$ ($n\leq N$). ($a_{8,1}\equiv a_8$ in section 2).
For the renormalized quantities in (29) we write, for example,
\def\N{^{(N)}}
\begin{equation}
E_i^{(N)}=(Z_8^{(N)})^{1/2}\left[E_i+Y_9^{(N)}\partial_iA_0+Y_{10}\N \partial_0A_i+Y_{11}\N A_0\wedge A_i+Y_{12}\N v_i \wedge c \right],
\end{equation}
%47
which has the same form as (29) but with $Y_9$ replaced by $Y_9\N$ etc, and similarly for the other fields and sources. Note that, in spite of the notation, because of the products,  $(Z_8^{(N)})^{1/2}Y_9^{(N)}$ etc, $E_i\N$  contains contributions higher than $O(\hbar^N)$, in fact up to $\hbar^{2N}$. The analogue of  (33) is
\begin{equation}
\tilde{\gamma}\N =\tilde{\gamma}_0\left(A_i\N,....,K\N ;g\N \right).
\end{equation}
%48
Just like $\tilde{\gamma}_R$ in (34), $\tilde{\gamma}\N$ satisfies the equation
\begin{equation}
\tilde{\gamma}\N * \tilde{\gamma}\N =0.
\end{equation}
%49
Again, in spite of the notation, $\tilde{\gamma}\N$ contains many terms of higher order than $\hbar^N$.
 One source of such terms is the analogue of the non-linear term (41). These higher order terms do not matter as long as we are working to $N$-loop order, but they are necessary for (49) to be exact, and are required
to proceed to higher orders.

Thus we make the assumption that (48) is the exact action we require, and then we get to 
 $(N+1)$-loop order in the same way as in section 3 equations (21) and (23). That is, we use
\def\M{^{(N+1)}}

\begin{equation}
G_{N+1}=a_{5,N+1} {u^i}\N .A_i^{(N)} +a_{6,N+1} u_0\N .A_0\N +a_{7,N+1} c\N .K\N $$
$$+a_{8,N+1} {v^i}\N.E_i\N 
+{v^i}\N .\big[a_{9,N+1}\partial_iA_0\N +a_{10,N+1}\partial_0 A_i\N $$ $$+a_{11,N+1} A_0\N \wedge A_i\N 
 + a_{12,N+1}  \frac{1}{2}(v_i\N  \wedge c\N )\big],
\end{equation}
where the coefficients $a_{5,N+1}$ etc are of order $\hbar^{N+1}$.
%50
Then the next order is obtained from
\begin{equation}
\tilde{\gamma}\N +\tilde{\gamma}\N *G_{N+1}.
\end{equation}
%51
In (50), we can replace $A_i\N$ by $A_i$ etc with neglect of terms $O(\hbar^{N+2})$.

To explain the argument, we show as an example the dependence on $E_i$; the other fields and sources
work out in analagous ways. We define
\begin{equation}
\Delta_{i,N+1}\equiv \frac{\delta G_{N+1}}{\delta v^i}$$
$$=a_{8,N+1}E_i+a_{9,N+1}\partial_iA_0+a_{10,N+1} \partial_0A_i+a_{11,N+1} A_0\wedge A_i+a_{12,N+1} v_i \wedge c .
\end{equation}
%52
Then (51) contains a contribution
\begin{equation}
\tilde{\gamma}\N +\int dx \left[\frac{\delta G\N}{\delta v^i(x)}\frac {\delta\tilde{\gamma}\N}{\delta E_i(x)}\right]+.....$$
$$\approx\tilde{\gamma}\N+\int dx \left[\Delta_{i,N+1} \frac {\delta\tilde{\gamma}\N}{\delta E_i(x)}\right]+.....$$
$$\approx\tilde{\gamma}\N(E_i+\Delta_{i,N+1},....)\approx \tilde{\gamma}(E_i\N+\Delta_{i,N+1},....)
\end{equation}
%53
with repeated neglect of terms of higher order than $\hbar^{N+1}$.
The argument of $\tilde{\gamma}$ in the last expression in (53) is, from (47) and (52).
\newpage
\begin{equation}
E_i\N+\Delta_{i,N+1}= $$
$$(Z_8^{(N)})^{1/2}\left[E_i+Y_9^{(N)}\partial_iA_0+Y_{10}\N \partial_0A_i+Y_{11}\N A_0\wedge A_i+Y_{12}\N v_i \wedge c \right]$$
$$+a_{8,N+1} E_i+a_{9,N+1}\partial_iA_0+a_{10,N+1} \partial_0A_i+a_{11,N+1} A_0\wedge A_i+a_{12,N+1} v_i \wedge c.
\end{equation}
%%54
We now use
\begin{equation}
(Z_8^{(N)})^{1/2}+a_{8,N+1}\approx (Z_8\M)^{1/2},
\end{equation}
%%55
\begin{equation}
(Z_8^{(N)})^{1/2}Y_9\N + a_{9,N+1}\approx (Z_8\M)^{1/2}[Y_9\N+a_{9,N+1}]=(Z_8\M)^{1/2}Y_9\M
\end{equation}
%%56
(again with neglect of $O(\hbar^{N+2}$), and similar equations involving $ a_{10}, a_{11}, a_{12}$.
Then, comparing with (47), (54) gives
\begin{equation}
E_i\N+\Delta_{i,N+1}=E_i\M + O(\hbar^{N+2}).
\end{equation}
Thus (51) gives (57), and so with neglect of higher order terms,
%57
\begin{equation}
\tilde{\gamma}(E_i\N+\Delta_{i,N+1},.....)\approx\tilde{\gamma}(E_i\M,.....)=\tilde{\gamma}\M,
\end{equation}
%58
as required for the iterative process to be consistent with (33).
\section{Summary}
We have studied the renormalization of a flow gauge, used to interpolate between the Feynman gauge
and the Coulomb gauge, thereby giving an unambiguous definition of the latter. The flow gauge has a parameter $\theta$. We use the Hamiltonian
formalism, which includes the momentum field $E_i^a$. The renormalized Lagrangian is obtained by mixing
fields and sources, as well as scaling (Section 2). We calculate the renormalization constants to one-loop order (section 3).
We find the general form of the renormalized Lagrangian (Section 4), and we demonstrate the order-by-order
renormalization is consistent with this (section 5). We find that counter-terms of a new structure appear first at two-loop order. The parameter $\theta$ has to be renormalized as well as the coupling constant.
\section{Appendix A: An example}
In this Appendix, we give a sample calculation of a divergent part which is typical of the calculations from which the counter-terms in (26) were derived. In particular, this example will show how the
denominators $1/(1+\t)^2$ arise. The amplitude which we choose is the $A_iA_0$ transition.
There are six  graphs, if we separate contributions from the transverse and longitudinal parts of the propagator in (8), but we treat only three of them here. These are shown in Fig.1.
\begin{figure}
\centering
	\includegraphics[width=0.90\textwidth]{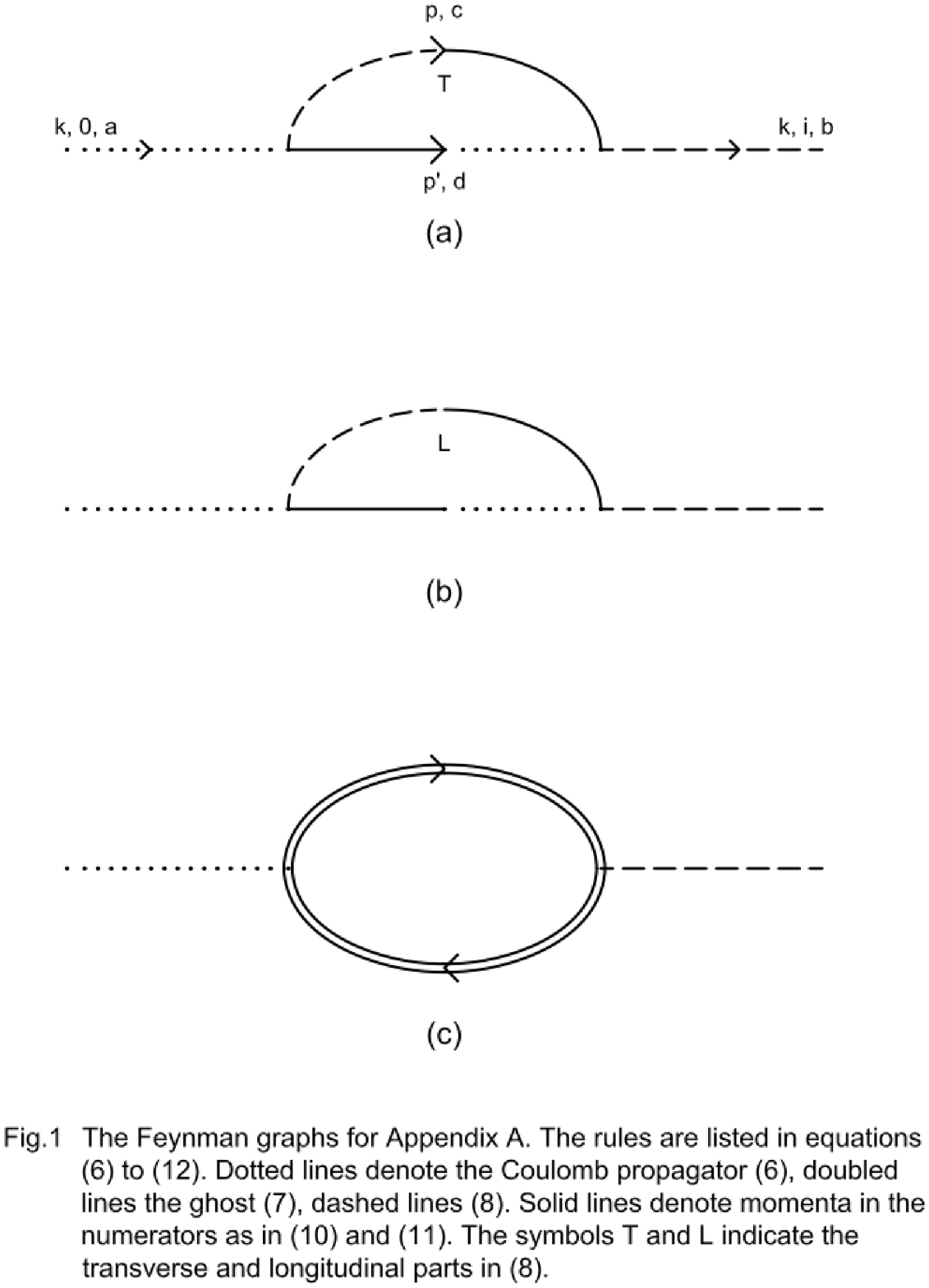}
	%	\caption{Feynman graphs For Appendix A. The rules are listed in equations (6) to (12). Dotted lines denote the Coulomb propagator (6), doubled lines lines the ghost (7), dashed lines (8).
	%Solid lines denote momenta in the numerators as in (10) and (11). The symbols $T$ and $L$ indicate the
	%transverse and longitudinal parts in (8).}
	\label{fig:Fig1}
\end{figure}

Graph (a) gives the integral
\begin{equation}
G^{a}=\z g^2C_G\delta_{ab}\int d^dp\frac{P'^j}{\bar{P'^2}}\frac{p_0}{p^2+i\eta}\left[g_j^i+\frac{P_jP^i}{P^2}\right].
\end{equation}
%59
Because of the transverse projection operator, $P'^j$ can be replaced by $K^j$. The integral over $p_0$
is zero for $k_0=0$, so a factor $k_0$ must come out. The integral is then only logarithmically
divergent, so to determine the divergent part it is sufficient to put $K=0$.
Then we require the  integral
\begin{equation}
I_1\equiv \int dp_0 \frac{p_0}{[p_0^2 -P^2][\t^2(p_0-k_0)^2-P^2]}.
\end{equation}
%60
To first order in $k_0$, this is
\begin{equation}
J_1k_0
\end{equation}
%61
where
\begin{equation}
J_1=\left[\frac{dI_1}{dk_0}\right]_{k_0=0} = 2\int dp_0\frac{\t^2 p_0^2}{[p_0^2-P^2][[\t^2p_0^2-P^2]^2}
=i\pi \frac{\t}{(1+\t)^2} \frac{1}{P^3}.
\end{equation}
%62
Inserting (62) and (61) into (59), we get
\begin{equation}
-\z g^2C_G \delta_{ab}K^jk_0\pi i\frac{\t}{(1+\t)^2}\int d^{d-1}P \left[g^i_j +\frac{P_jP^i}{P^2}\right]\frac{1}{P^3}$$
$$=-\z g^2C_G \delta_{ab}K^ik_0\pi i\frac{\t}{(1+\t)^2} \frac{2}{3}\frac{4\pi}{\epsilon},
\end{equation}
%63
where, in dimensional regularization, $d=4-\epsilon$ and (63) is just the pole part of (59).

It is convenient to take the contributions from graphs (b) and (c) of Fig.1 together. They give
\begin{equation}
G^{b}+G^{c}=\z g^2C_G\delta_{ab}\t^2\int d^dp\frac{p_0}{\bar{P^2}\bar{P'^2}}\left[\frac{P^iP_l}{P^2}+g^i_l\right]P'^l,
\end{equation}
%64
where the first term in the square bracket comes from graph (b) and the second from (c).
Again the transverse projection operator allows us to replace $P'^l$ by $K^l$.
The $p_0$ integral is
\begin{equation}
I_2\equiv \int dp_0 \frac{p_0}{(\t^2p_0^2 -P^2)[\t^2(p_0-k_0)^2-P^2]},
\end{equation}
%65
which to first order in $k_0$ gives $J_2k_0$, where
\begin{equation}
J_2=\int dp_0\frac{2\t^2p_0^2}{(\t^2p_0^2-P^2)^3}=\frac{i\pi}{4\t}\frac{1}{P^3}.
\end{equation}
%66
Inserting (65) and (66) into (64), and integrating over $P$ as in (63), we get
\begin{equation}
\z g^2C_G \delta_{ab}K^ik_0\pi i\t \frac{1}{6}\frac{4\pi}{\epsilon}.
\end{equation}
%67
Combining (68) with (71) and expressing the result of the divergent coefficient $c$ defined in (25),
the total result for graphs (a), (b) and (c) is
\begin{equation}
ic\t\frac{(\t^2+2\t-3)}{(1+\t)^2}\delta_{ab}k_0K^i.
\end{equation}
%68
\begin{figure}
\centering
	\includegraphics[width=0.90\textwidth]{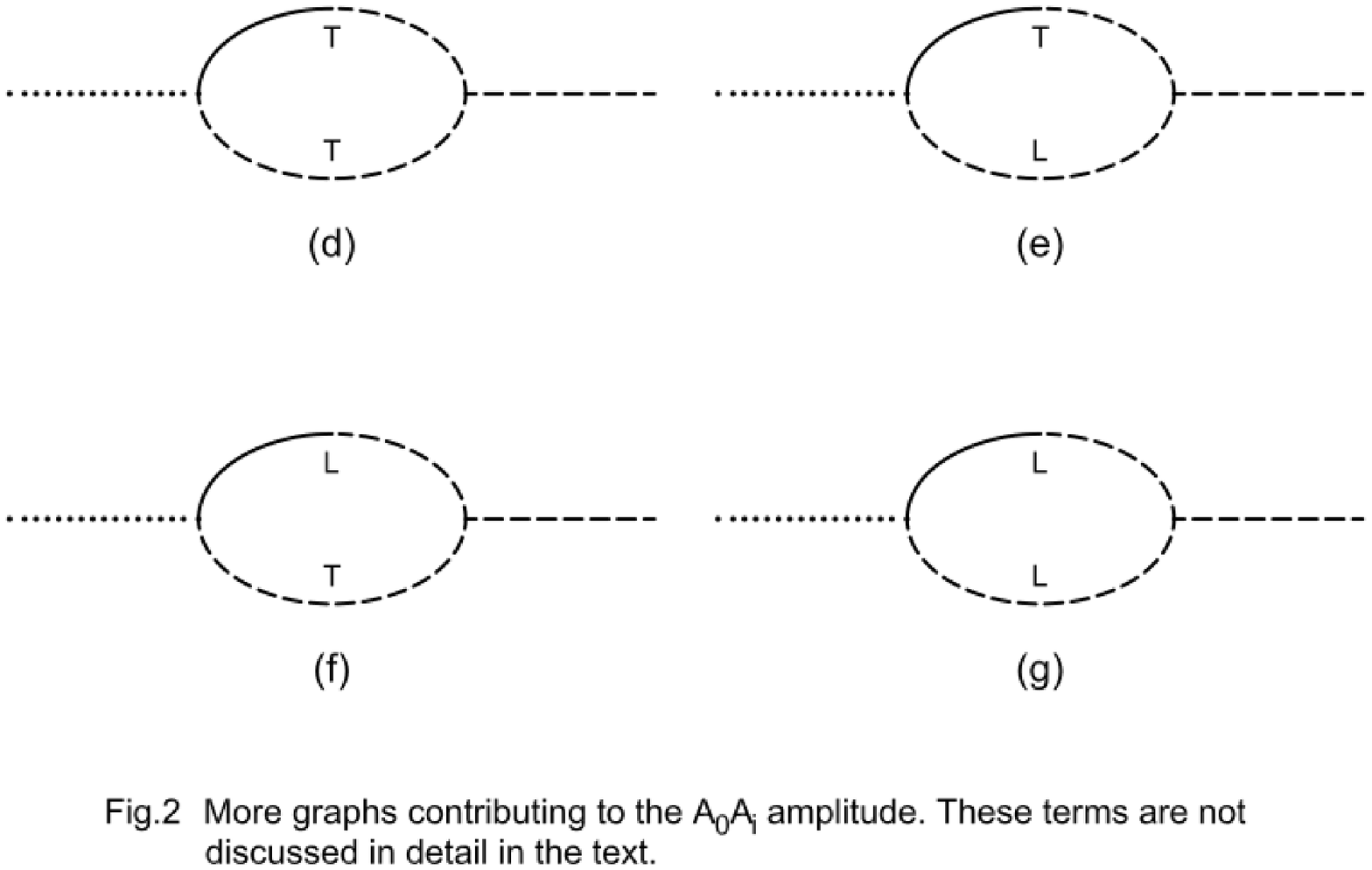}
	%	\caption{Feynman graphs For Appendix A. The rules are listed in equations (6) to (12). Dotted lines denote the Coulomb propagator (6), doubled lines lines the ghost (7), dashed lines (8).
	%Solid lines denote momenta in the numerators as in (10) and (11). The symbols $T$ and $L$ indicate the
	%transverse and longitudinal parts in (8).}
	\label{fig:Fig2}
\end{figure}
The graphs in Fig.2 contain the right hand vertex
\begin{equation}
igf^{bdc}[g^{jl}(P-P')^i+g^{li}(K+P')^j-g^{ij}(P+K)^l]\equiv gf^{bdc}V^{ijl}.
\end{equation}
%69
The integrand is simplified by the use of the identity
\begin{equation}
P_jV^{ijl}=-(K^2g^{il}+K^iK^l)+(P'^2g^{il}+P'^iP'^l),
\end{equation}
%70
where the first term leads to a convergent integral, and the second term acts as a projection operator as in (59) and (64).
We just quote the results for the graphs in Fig.2. Graph (d) gives
\begin{equation}
ic\frac{1}{3}\delta_{ab}k_0K^i,
\end{equation}
%71
and (e) and (f) together give
\begin{equation}
-ic\t^2\frac{4}{3}\frac{(1-\t)}{(1+\t)^2}\delta_{ab}k_0K^i.
\end{equation}
%72

Graph (g) is not divergent. The complete divergent part for the graphs of Fig.1 and Fig.2 is
\begin{equation}
ic\frac{1}{3}\left[1-\t\frac{(1-\t)(3+5\t)}{(1+\t)^2}\right]\delta_{ab}k_0K^i.
\end{equation}
%73
This is canceled by the counter-term proportional to $(a_9-a_{10})$ in the second line of  (24).

\section{Appendix B: Theorem}
Here is a proof that the anticanonical conditions (35) and (36) imply the BRST property (34). Our argument is not completely general but is sufficient for the present case.

We use the highly abbreviated notation
\def\a{\alpha}
\def\b{\beta}
\def\D{\delta}
\begin{equation} \psi^\a =\{A^{ia}(x), A_0^b(x), E^{jc}(x), K^d(x)\},$$
$$ \chi_\b=\{u_i^a(x),u_0^b(x), v_j^c(x), c^d(x)\}
\end{equation}
%74
where the indices $\a$ and $\b$ stand for all the vector and colour indices and for $x$.
Repeated indices are summed over, which includes integration over $x$. The above definition has been chosen so that $\psi$ consists of bosonic variables  and $\chi$ consists of fermionic variables.
This entails that the source $K$ is in $\psi$ and the field $c$ is in $\chi$.

With this notation, the BRST operation (13) becomes
\begin{equation}
X*Y=\frac{\D X}{\D \chi_\a}\frac{\D Y}{\D \psi^\a}\pm \frac{\D Y}{\D \chi_\a}\frac{\D X}{\D \psi^\a},
\end{equation}
%75
where the minus sign applies if $X,Y$ are both fermionic and the plus sign otherwise.
(This is the same ordering as we used in (13), with fermionic denominators on the left.)
Then
\begin{equation}
\psi^\a * \chi_\b =\delta^\a_\b,\,\,\,\, \psi^\a * \psi^\b=0, \,\,\,\,\, \chi_\a * \chi_\b=0.
\end{equation}
%76

For the renormalized quantities, we use the notation
\begin{equation} \psi^\a_R =\{A^{ia}_R(x), A_{R0}^b(x), E^{jc}_R(x), K^d_R(x)\},$$
$$ \chi_{R\b}=\{u_{Ri}^a(x),u_{R0}^b(x), v_{Rj}^c(x), c_R^d(x)\}.
\end{equation}
%77
Then the conditions (35) and (36) for the transformation  to the renormalized quantities to be anticanonical are
\begin{equation}
\psi_R^\a * \chi_{R\b} =\delta^\a_\b,\,\,\,\, \psi_R^\a * \psi_R^\b=0, \,\,\,\,\, \chi_{R\a} * \chi_{R\b}=0.
\end{equation}
%78
\def\g{\gamma}
Using (75), these conditions become
\begin{equation}
\left[\frac{\D\psi_R^\a}{\D\chi_\g}\frac{\D\chi_{R\b}}{\D\psi^\g}+\frac{\D\chi_{R\b}}{\D\chi_\g}\frac{\D\psi_R^\a}{\D\psi^\g}\right]=\delta^\a_\b,
\end{equation}
%79
\begin{equation}
\left[\frac{\D\psi_R^\a}{\D\chi_\g}\frac{\D\psi_R^\b}{\D\psi^\g}+ (\a\leftrightarrow \b)\right]=0,\,\,\,\,
\left[\frac{\D\chi_{R\a}}{\D\chi_\g}\frac{\D\chi_{R\b}}{\D\psi^\g}- (\a\leftrightarrow \b)\right]=0.
\end{equation}
%80
Since $\tilde{\gamma}_R$ is the same function of ($\chi_{R\a},\psi_R^\b$) as $\tilde{\gamma}$ is of ($\chi_\a,\psi^\b$),  equation (74) and the property $\tilde{\gamma}_0 *\tilde{\gamma}_0=0$ imply that
\begin{equation}
\frac{\D \tilde{\gamma}_R}{\D \chi_{R\a}}\frac{\D \tilde{\gamma}_R}{\D {\psi_R}^\a}=0.
\end{equation}
%81
\def\G{\tilde{\gamma}}

With these preliminaries, we can complete the proof of the theorem.
\begin{equation}
\frac{1}{2}\tilde{\gamma}_R * \G_R=\frac{\D\G_R}{\D\chi_\a}\frac{\D\G_R}{\D\psi^\a}=
\left[\frac{\D\G_R}{\D\psi_R^\b}\frac{\D\psi_R^\b}{\D\chi_\a}+\frac{\D\G_R}{\D\chi_{R\b}}\frac{\D\chi_{R\b}}{\D\chi_\a}\right] \left[\frac{\D \psi_R^\g}{\D\psi^\a}\frac{\D\tilde{\gamma}_R}{\D\psi_R^\g}+
\frac{\D\chi_{R\g}}{\D\psi^\a}\frac{\D\tilde{\gamma}_R}{\D\chi_{R\g}}\right]
\end{equation}
%82
%We will need the chain rules for differentiation, taking account of anti-commutation relations.
%\begin{equation}
%\frac{\D \tilde{\gamma}_R}{\D\psi^\a}=\frac{\D \phi_R^\b}{\D\psi^\a}\frac{\D\tilde{\gamma}_R}{\D\phi_R^\b}+\frac{\D\chi_R^\b}{\D\psi^\a}\frac{\D\tilde{\gamma}_R}{\D\chi_R^\b},\,\,\,\,\frac{\D \tilde{\gamma}_R}{\D\chi_\a}=
%\frac{\D \phi_R^\b}{\D\chi_\a}\frac{\D\tilde{\gamma}_R}{\D\phi_R^\b}+\frac{\D\chi_R^\b}{\D\chi_\a}\frac{\D\tilde{\gamma}_R}{\D\chi_R^\b}.
%\end{equation}
Here the only question about ordering occurs in the product $\frac{\D\chi_{R\g}}{\D\psi^\a}\frac{\D\tilde{\gamma}_R}{\D\chi_{R\g}}$ where both factors are fermionic.
The above ordering is correct because in
$\tilde{\gamma}_R$ the  terms in $\chi_{R\a}$ which contribute are functional derivatives   with respect to $A_i$ or $A_0$ of $u_{R0}$ and $u_{Ri}$, and these come on the left in $\tilde{\gamma}_R$ in (33) just as $u_0$ and $u_i$  are on the left in $\tilde{\gamma}_0$ in (3).
In the other terms in  (82), the ordering is unimportant. 
We may write (82) as the sum of two parts:
\begin{equation}
\frac{\D\G_R}{\D\psi_R^\b}\frac{\D\psi_R^\b}{\D\chi_\a}\frac{\D\chi_{R\g}}{\D\psi^\a}\frac{\D \G_R}{\D\chi_{R\g}}
+\frac{\D\G_R}{\D\chi_{R\b}}\frac{\D\chi_{R\b}}{\D\chi_\a}\frac{\D\psi_R^\g}{\D\psi^\a}\frac{\D \G_R}{\D\psi_R^\g}
\end{equation}
%83
\begin{equation}
\frac{\D\G_R}{\D\psi_R^\b}\frac{\D\psi_R^\b}{\D\chi_\a}\frac{\D\psi_R^\g}{\D\psi^\a}
\frac{\D \G_R}{\D\psi_R^\g} + \frac{\D\G_R}{\D\chi_{R\b}}\frac{\D\chi_{R\b}}{\D\chi_\a}\frac{\D\chi_{R\g}}{\D\psi^\a}
\frac{\D \G_R}{\D\chi_{R\g}}.
\end{equation}
%84
In the second term in (83)  only the first factor is fermionic and (83) can be re-ordered to give (interchanging also $\b$ and $\g$)
\begin{equation}
\frac{\D\G_R}{\D\psi_R^\b}\left[\frac{\D\psi_R^\b}{\D\chi_\a}\frac{\D\chi_{R\g}}{\D\psi^\a}
+\frac{\D\chi_{R\g}}{\D\chi_\a}\frac{\D\psi_R^\b}{\D\psi^\a}\right]\frac{\D \G_R}{\D\chi_{R\b}}=0,
\end{equation}
%85
using (79) and (81).
Similarly, each term in (84) is zero by (80). This completes the proof that (82) is zero.

\end{document}